\documentclass[onecolumn,11pt,draft=false]{IEEEtran}
\usepackage{amsthm, amssymb, amsmath}

\usepackage[bookmarks, colorlinks=true, citecolor = blue]{hyperref}
\usepackage{booktabs}
\usepackage{longtable}
\usepackage[font=small]{caption}
\usepackage{pdfsync}
\usepackage{mathtools}
    
\renewcommand{\baselinestretch}{1.58}

\begin{document}
\title{\bf\LARGE R\'{e}nyi Cross-Entropy Measures for Common Distributions \\ and Processes with Memory
}

\author{
Ferenc Cole Thierrin,   
Fady Alajaji, Tam\'{a}s Linder \\ \smallskip
\small Department of Mathematics and Statistics \\ Queen's University \\Kingston, ON K7L 3N6, Canada \\ 
Emails: \{14fngt, fa, tamas.linder\}@queensu.ca
}
\maketitle
\begin{abstract}
Two R\'{e}nyi-type generalizations of the Shannon cross-entropy, the R\'{e}nyi cross-entropy and the Natural R\'{e}nyi cross-entropy, were recently used as loss functions for the improved design of deep learning generative adversarial networks. In this work, we build upon our results in \cite{cole} by deriving the R\'{e}nyi and Natural R\'{e}nyi differential cross-entropy measures in closed form for a wide class of common continuous distributions belonging to the exponential family and tabulating the results for ease of reference. We also summarise the R\'{e}nyi-type cross-entropy rates between stationary Gaussian processes and between finite-alphabet time-invariant Markov sources.
\end{abstract}
\begin{center} 
{\bf Keywords}
\end{center}
\vspace{-0.05in}
R\'{e}nyi information measures, cross-entropy, divergence measures, exponential family distributions, Gaussian processes, Markov sources. 



\bigskip\bigskip

\section{Introduction}
The R\'{e}nyi entropy \cite{renyi} of order $\alpha$ of a probability mass function $p$ with finite support $\mathbb{S}$ is defined as $$H_\alpha(p)=\frac{1}{1-\alpha}\ln \sum_{x\in\mathbb{S}}p(x)^\alpha $$ for $\alpha>0,\alpha\neq 1$. The R\'{e}nyi entropy generalizes the Shannon entropy, $$H(p)=-\sum_{x\in \mathbb{S}} p(x)\ln p(x),$$ in the sense that as $\alpha \to 1$, $H_\alpha(p) \to H(p)$. Several other R\'{e}nyi-type information measures have been put forward, each obeying the condition that their limit as $\alpha$ goes to one reduces to a Shannon-type information measure. This includes the R\'{e}nyi divergence (of order $\alpha$) between two discrete distributions $p$ and $q$ with common finite support $\mathbb{S}$, given by $$D_\alpha(p||q)= \frac{1}{\alpha-1}\ln \sum_{x\in\mathbb{S}} p(x)^\alpha q(x)^{1-\alpha}$$
which reduces to the familiar Kullback-Leibler divergence, 
$$D_{\text{\tiny KL}}(p||q)=  \sum_{x\in\mathbb{S}} p(x) 
\ln \frac{p(x)}{q(x)},$$
as $\alpha \to 1$. Note that in some cases \cite{mutualInfo}, there may exist multiple R\'{e}nyi-type generalisations for the same information measure (particularly for the mutual information).

Many of these definitions admit natural counterparts in the case when the involved distributions have a probability density function (pdf). This gives rise to information measures such as the R\'{e}nyi differential entropy for pdf $p$ with support $\mathbb{S}$, $$h_\alpha(p)=\frac{1}{1-\alpha}\ln \int_\mathbb{S}p(x)^\alpha \, dx,$$ and the R\'{e}nyi (differential) divergence between pdfs $p$ and $q$ with common support $\mathbb{S}$, $$D_\alpha(p\|q)= \frac{1}{\alpha-1}\ln \int_\mathbb{S} p(x)^\alpha q(x)^{1-\alpha} \, dx.$$

The R\'{e}nyi cross-entropy between distributions $p$ and $q$ is an analogous generalization of the Shannon cross-entropy $$H(p;q)=-\sum_{x \in \mathbb{S}} p(x)\ln q(x).$$ Two definitions for this measure have been recently suggested. In light of the fact that Shannon's cross-entropy satisfies $H(p;q)=D(p\|q)+H(p)$, a natural definition of the R\'{e}nyi cross-entropy is: 
\begin{equation}\label{rgan-ce} \tilde H_\alpha(p;q)\coloneqq D_\alpha(p||q)+H_\alpha (p).
\end{equation}
This definition was indeed proposed in~\cite{rgan} in the continuous case, with the differential cross-entropy measure given by
\begin{equation}\label{rgan-ce2} \tilde h_\alpha(p;q)\coloneqq D_\alpha(p||q)+h_\alpha (p).
\end{equation}
In contrast, prior to \cite{rgan}, the authors of \cite{Alba} introduced the R\'{e}nyi cross-entropy in their study of shifted R\'{e}nyi measures expressed as the logarithm of weighted generalized power means. Specifically, upon simplifying Definition~6 in~\cite{Alba}, their expression for the R\'{e}nyi cross-entropy between distributions $p$ and $q$ is  given by 
\begin{equation}\label{alba-ce}
H_\alpha\left(p;q\right)\coloneqq\frac{1}{1-\alpha}\ln\sum_{x\in\mathbb{S}} p\left(x\right)q\left(x\right)^{\alpha-1}.    
\end{equation}
For the continuous case, \eqref{alba-ce} can be readily converted to yield the R\'{e}nyi differential cross-entropy between pdfs $p$ and $q$:
\begin{equation} \label{renyi-diff-ce}
h_\alpha(p;q)\coloneqq\frac{1}{1-\alpha}\ln \int_\mathbb{S} p(x)q(x)^{\alpha-1} \, dx.
\end{equation}

Note that both  \eqref{rgan-ce} and \eqref{alba-ce} reduce to the Shannon cross-entropy $H(p;q)$ as $\alpha\to 1$ \cite{Bhatia}. A similar result holds for \eqref{rgan-ce2} and \eqref{renyi-diff-ce}, where the Shannon differential cross-entropy, $h(p;q) = -\int_\mathbb{S} p(x) \ln q(x) \, dx$, is obtained.
Also, the R\'{e}nyi (differential) entropy is recovered in all equations when $p=q$ (almost everywhere). These properties alone make these definitions viable extensions of the Shannon (differential) cross-entropy. 

Finding closed-form expressions for the cross-entropy measure in \eqref{rgan-ce2} for continuous distributions is direct, since the R\'{e}nyi divergence and the R\'{e}nyi differential entropy were already calculated for numerous distributions in \cite{Gil} and \cite{Song}, respectively. However, deriving the measure in \eqref{renyi-diff-ce} is more involved. We hereafter refer to the measures $\tilde H_\alpha(p;q)$ in~\eqref{rgan-ce} and $\tilde h_\alpha(p;q)$~\eqref{rgan-ce2}, as the \textit{Natural R\'{e}nyi cross-entropy} and the \textit{Natural R\'{e}nyi differential cross-entropy}, respectively; while we plainly call the measures in $H_\alpha(p;q)$ in \eqref{alba-ce} and $h_\alpha(p;q)$ \eqref{renyi-diff-ce} as the \textit{R\'{e}nyi cross-entropy} and the \textit{R\'{e}nyi differential cross-entropy}, respectively.

In a recent conference paper \cite{cole}, we showed how to calculate the R\'{e}nyi differential cross-entropy $h_\alpha(p;q)$ between distributions of the same type from the exponential family. Building upon the results shown there, the purpose of this paper is to derive in closed form the expression of $h_\alpha(p;q)$ for thirteen commonly used univariate distributions from the exponential family, as well as for multivariate Gaussians, and tabulate the results for ease of reference. We also analytically derive the Natural R\'{e}nyi differential cross-entropy $\tilde h_\alpha(p;q)$ for the same set of distributions.
Finally, we present tables summarising the R\'{e}nyi and Natural R\'{e}nyi (differential) cross-entropy rate measures, along with their Shannon counterparts, for two important classes of sources with memory, namely stationary Gaussian sources and finite-state time-invariant Markov sources.

Motivation for determining formulae for the R\'{e}nyi cross-entropy originates from the use of the Shannon differential cross-entropy as a loss function for the design of deep learning generative adversarial networks (GANs) in \cite{Goodfellow2014}. The parameter $\alpha$, ubiquitous to all R\'{e}nyi information measures, allows one to fine-tune the loss function to improve the quality of the GAN-generated output. This can be seen in \cite{paper,Bhatia} and \cite{rgan}, which used the R\'{e}nyi differential cross-entropy, and the Natural R\'{e}nyi differential cross-entropy measures, respectively, to generalize the original GAN loss function (which is recovered as $\alpha \to 1$), resulting in both improved GAN system stability and performance for multiple image datasets. It is also shown in \cite{paper,Bhatia} that the introduced R\'{e}nyi-centric generalized loss function
preserves the equilibrium point satisfied by the original GAN via the so-called Jensen-R\'{e}nyi divergence \cite{kluza19}, a natural extension of the Jensen-Shannon divergence \cite{jsd} upon which the equilibrium result of \cite{Goodfellow2014} is established. Other GAN systems that utilize different generalized loss functions were recently developed and analyzed in~\cite{cumulant_gan} and \cite{kurri2021realizing, kurri2022} (see also the references therein for prior work).

The rest of this paper is organised as follows. In Section~\ref{sec2}, the formulae for the R\'{e}nyi differential cross-entropy and Natural R\'{e}nyi differential cross-entropy for distributions from the exponential family are given. In Section~\ref{sec3}, these calculations are systematically carried for fourteen pairs of distributions of the same type within the exponential family and the results are presented in two tables. The R\'{e}nyi and Natural R\'{e}nyi differential cross-entropy rates are presented in Section~\ref{sec4} for stationary Gaussian sources; furthermore, the R\'{e}nyi and Natural R\'{e}nyi cross-entropy rates are provided in Section~\ref{sec5} for finite-state time-invariant Markov sources. Finally, the paper is concluded in Section~\ref{sec6}.

\section{R\'{e}nyi and Natural R\'{e}nyi Differential Cross-Entropies for Distributions from the Exponential Family}
\label{sec2}
An exponential family is a class of probability distributions over a support $\mathbb{S} \subseteq \mathbb{R}^n$ defined by a parameter space $\Theta \subseteq \mathbb{R}^m$ and functions $b: \mathbb{S} \mapsto \mathbb{R}$, $c: \Theta \mapsto \mathbb{R}$, $T: \mathbb{S} \mapsto \mathbb{R}^m$, and $\eta: \Theta \mapsto \mathbb{R}^m$ such that the pdf in this family have the form 
\begin{equation} \label{exp-f1}
f(x)=c (\theta)b (x)\exp \left(\langle \eta(\theta), T (x)\rangle \right), \qquad x \in \mathbb{S}
\end{equation} 
where $\langle\cdot,\cdot\rangle$ denotes the standard inner product in $\mathbb{R}^m$.
Alternatively, by using the (natural) parameter $\eta=\eta(\theta)$, the pdf can also be written
\begin{equation} \label{exp-f2}
    \displaystyle{f(x) =b (x)\exp \left(\langle \eta, T (x)\rangle +A(\eta)\right)}, 
    \end{equation}
 where $A(\eta): \eta(\Theta) \mapsto \mathbb{R}$ with $A(\eta)=\ln c(\theta)$.
 Examples of important pdfs we consider from the exponential family are included in Appendix A.
 
 In \cite{cole} the cross-entropy between pdfs $f_1$ and $f_2$ of the same type from the exponential family was proven to be
\begin{equation}
h_\alpha\left(f_1;f_2\right)=\frac{A\left(\eta_1\right)-A\left(\eta_h\right)+\ln E_h}{1-\alpha}-A\left(\eta_2\right), \label{eq:three}
\end{equation}
where $$E_h= \mathbb{E}_{f_h}\left[b(X)^{\alpha-1}\right]=\int b(x)^{\alpha-1}f_h(x) \, dx.$$ Here $f_h$ refers to a distribution of the same type as $f_1$ and $f_2$ within the exponential family with natural parameter $$\eta_h \coloneqq \eta_1+(\alpha-1)\eta_2.$$

It can also be shown that the Natural R\'{e}nyi differential cross-entropy between $f_1$ and $f_2$ is given by
\begin{equation}
    \tilde h_\alpha\left(f_1;f_2\right)=\frac{A\left(\eta_\alpha\right)-A\left(\alpha \eta_1\right)+\ln E_\alpha}{1-\alpha}-A\left(\eta_2\right),\label{eq:nat}
\end{equation}
where 
$$\eta_\alpha=\alpha\eta_1+(1-\alpha)\eta_2,$$
and
$$E_\alpha = \mathbb{E}_{f_{\alpha 1}}\left[b(X)^{\alpha-1}\right]=\int b(x)^{\alpha-1}f_{\alpha 1}(x) \, dx$$ 
where $f_{\alpha 1}$ refers to a distribution of the same type as $f_1$ and $f_2$ within the exponential family with natural parameter $\alpha\eta_1$. 

\section{Tables of R\'{e}nyi and Natural R\'{e}nyi Differential Cross-Entropies} \label{sec3}
Tables \ref{RenDivTable} and \ref{RenNatTable} list R\'{e}nyi and Natural differential cross-entropy expressions, respectively, between common distributions of the same type from the exponential family (which we describe in Appendix~A for convenience). 
The closed-form expressions were derived using \eqref{eq:three} and \eqref{eq:nat}, respectively.
In the tables, the subscript of $i$ is used to denote that a parameter belongs to pdf $f_i$, $i=1,2$. 


\bigskip

\begin{longtable}[hbtp]{cc}\caption{R\'{e}nyi Differential Cross-Entropies}\label{RenDivTable} 

\\
\hline
\textbf{Name} & $h_\alpha(f_1;f_2)$ \\ \hline
\hline

\textbf{Beta} & $\displaystyle{\ln {B(a_2, b_2)} +  \frac{1}{\alpha - 1}\ln \frac{B(a_h, b_h)}{B(a_1, b_1)}} $\\\cline{2-2} 
& $a_h \coloneqq a_1 + (\alpha-1)(a_2-1)$,$\hspace{1cm} a_h > 0$\\& $b_h \coloneqq b_1 + (\alpha-1)(b_2-1)$,$\hspace{1cm} b_h > 0$ \\ 

\hline
$\begin{array}{c}
\boldsymbol{\chi}\\ \textbf{ (scaled)}\end{array}$ & $\displaystyle{\frac{1}{2}\left(k_2\ln\sigma_2^2\sigma_h^2-\ln2\sigma_h^2\right) +\ln\Gamma\left(\frac{k_2}{2}\right)}$\\
&$+ \displaystyle{ \frac{1}{\alpha - 1}\left( \ln \Gamma\left(\frac{k_h}{2}\right) -\ln \Gamma\left(\frac{k_1}{2}\right)-\frac{k_1}{2}\ln\sigma_1^2\sigma_h^2 \right)}$  \\\cline{2-2} 
	& $\sigma^{2}_h \coloneqq \frac{1}{\sigma_1^2}+ \frac{\alpha-1}{\sigma_2^2}$,$\hspace{1cm} \sigma^2_h > 0$\\
	& $k_h \coloneqq k_1+(\alpha-1)(k_2-1)$,$\hspace{1cm} k_h > 0$\\

\hline 
$\begin{array}{c}
\boldsymbol{\chi}\\ \textbf{ (non-scaled)}\end{array}$ & $\displaystyle{\frac{1}{2}\left(k_2\ln\alpha-\ln2\alpha\right) +\ln\Gamma\left(\frac{k_2}{2}\right)}$\\
&$+ \displaystyle{ \frac{1}{\alpha - 1}\left( \ln \Gamma\left(\frac{k_h}{2}\right) -\ln \Gamma\left(\frac{k_1}{2}\right)-\frac{k_1}{2}\ln\alpha \right)}$  \\\cline{2-2} 
	& $k_h \coloneqq k_1+(\alpha-1)(k_2-1)$,$\hspace{1cm} k_h > 0$\\

\hline 
$\boldsymbol{\chi^2}$ & $\displaystyle{\frac{1}{1-\alpha}\left( \frac{\nu_1}{2}\ln\left(\alpha\right)-\ln\Gamma\left(\frac{\nu_1}{2}\right)+\ln\Gamma\left(\frac{\nu_h}{2}\right)\right)}$\\
&$+\displaystyle{\frac{2-\nu_2}{2}\ln\left(\alpha \right)+\ln2\Gamma\left(\frac{\nu_2}{2}\right)} $  \\\cline{2-2} 
	& $\nu_h \coloneqq \nu_1 + (\alpha-1)(\nu_2-2)$,$\hspace{1cm} \nu_h > 0$\\

\hline

\textbf{Exponential} & $\displaystyle{\frac{1}{1-\alpha}\ln \frac{\lambda_1}{\lambda_h} -\ln \lambda_2 }$  \\\cline{2-2} 
& $\lambda_h \coloneqq \lambda_1 + (\alpha-1)\lambda_2$,$\hspace{1cm} \lambda_h > 0$ \\

\hline 
\textbf{Gamma} & $\displaystyle{\ln \Gamma(k_2)+ k_2\ln\theta_2}$\\
& $+ \displaystyle{ \frac{1}{1-\alpha}\left( \ln\frac{\Gamma(k_h)}{\Gamma(k_1)}-k_h\ln\theta_h-k_1\ln\theta_1 \right) } $ \\\cline{2-2} 
&  $\theta_h \coloneqq \frac{\theta_1 + (a-1)\theta_2}{(\alpha-1)\theta_1\theta_1}$,$\hspace{1cm} \theta_h > 0$\\  &$k_h \coloneqq k_1 + (\alpha-1)k_2$,$\hspace{1cm} k_h > 0$  \\ 
\hline 
\pagebreak
\hline
\hspace{-8pt}$\begin{array}{c}
\textbf{Gaussian}\\ \textbf{ (univariate)}\end{array}$
 & $\displaystyle{\frac{1}{2}\left(\ln (2\pi\sigma_2^2) +  \frac{1}{1-\alpha}\ln \left( \frac{\sigma_2^2}{(\sigma^2)_h}\right) + \frac{(\mu_1 - \mu_2)^2}{ (\sigma^2)_h} \right)} $\\\cline{2-2} 
& $(\sigma^2)_h \coloneqq \sigma_2^2 + (\alpha-1)\sigma_1^2$,$\hspace{1cm} (\sigma^2)_h > 0$ \\
\hline
\hline
$\begin{array}{c}
\textbf{Gaussian}\\ \textbf{ (Multivariate)}\end{array}$
 & $\displaystyle{\frac{1}{2-2\alpha}\big(-\ln|A||\Sigma_1|+\left(1-\alpha\right)\ln\left(2\pi\right)^n|\Sigma_2|-d\big) } $\\\cline{2-2} 
& $A \coloneqq \Sigma_1^{-1}+(\alpha-1)\Sigma_2^{-1}$ , $\hspace{1cm} A \succ 0$\\
&$\displaystyle{d\coloneqq\boldsymbol{\mu^T _1}\Sigma^{-1} _1 \boldsymbol{\mu_1} + \left(\alpha-1\right)\boldsymbol{\mu^T _2}\Sigma^{-1} _2\boldsymbol{\mu_2}}$\\&$\displaystyle{-(\boldsymbol{\mu^T _1}\Sigma^{-1} _1  + \left(\alpha-1\right)\boldsymbol{\mu^T _2}\Sigma_2^{-1}) A^{-1}(\Sigma^{-1} _1 \boldsymbol{\mu_1} + \left(\alpha-1\right)\Sigma^{-1} _2 \boldsymbol{\mu_2})}$\\
\hline
\hspace{-8pt}
$\begin{array}{c}
\textbf{Gumbel}\\
\boldsymbol{(\beta_1 = \beta_2 =\beta)}\
\end{array}$ & $\displaystyle{ \frac{1}{1-\alpha}\left(\ln \frac{ \Gamma(2-\alpha)}{\beta}-\frac{\mu_1}{\beta}-\alpha\ln \eta_h\right)+\frac{\mu_2}{\beta}}$ \\\cline{2-2} 
& $\displaystyle{\eta_h\coloneqq e^{-\mu_1/\beta}+(\alpha-1)e^{-\mu_2/\beta}}$,$\hspace{1cm} \eta_h > 0$\\

\hline
%
\textbf{Half-Normal}  & $\displaystyle{\frac{1}{2}\left(\ln (\frac{\pi\sigma_2^2}{2}) +  \frac{1}{1-\alpha}\ln \left( \frac{\sigma_2^2}{(\sigma^2)_h}\right) \right)} $\\\cline{2-2} 
& $(\sigma^2)_h \coloneqq \sigma_2^{2} + (\alpha-1)\sigma_1^2$,$\hspace{1cm} (\sigma^2)_h > 0$ \\
\hline
$\begin{array}{c}
\textbf{Laplace}\\
\textbf{$(\mu_1=\mu_2=0)$}\\
\end{array}$
& $\displaystyle{\ln (2 b_2) +  \frac{1}{1-\alpha}\ln\left(\frac{b_2}{2b_h}\right)}$
\\\cline{2-2} 
 & $\displaystyle{b_h \coloneqq b_2+(1-\alpha)b_1}$,$\hspace{1cm} b_h > 0$\\
\hline
\hspace{-8pt}
$\begin{array}{c}
\textbf{Maxwell}\\
\textbf{\textbf{Boltzmann}}\\
\end{array}$ 
&$\displaystyle{\frac{1}{2}\left(\ln 2\pi +3\ln \sigma_2^2 \right)+\ln\sigma_h^2}$  \\ & $+\displaystyle{\frac{1}{1-\alpha}\left(\ln  \frac{\Gamma(2\alpha)}{\Gamma(\alpha)} -\frac{3}{2}\ln\sigma_1^2\sigma_h^2 \right) }$\\\cline{2-2} 
& $\sigma^2_h \coloneqq \sigma_1^{-2} + (\alpha-1)\sigma_2^{-2}$,$\hspace{1cm} \sigma^2_h > 0$ \\

\hline
$\begin{array}{c}
\textbf{Pareto}\\
\boldsymbol{(m_1 = m_2=m)}
\end{array}$ 
& $\displaystyle{- \ln m - \ln  \lambda_2 +  \frac{1}{1-\alpha}\ln \frac{\lambda_1}{\lambda_h}}$
 \\\cline{2-2} 
& $\lambda_h\coloneqq\lambda_1+\left(\alpha-1\right)\left(\lambda_2+1\right)$,$\hspace{1cm} \lambda_h > 0$ \\

\hline
\textbf{Rayleigh} & $\displaystyle{\frac{\ln \sigma_1^2-\alpha\ln\sigma_h^2+\ln\Gamma(\frac{1-\alpha}{2})}{1-\alpha}+\ln 2\sigma_2^2}
$ \\\cline{2-2} 
& $\sigma_h^2\coloneqq\sigma_1^{-2}+(\alpha-1)\sigma_2^{-2}$,$\hspace{1cm} \sigma^2_h > 0$ \\ 
\hline 
\end{longtable}

\newpage

\begin{longtable}[hbtp]{cc}\caption{Natural R\'{e}nyi Differential Cross-Entropies}\label{RenNatTable} 
\\
\hline
\textbf{Name} & $\tilde h_\alpha(f_1;f_2)$ \\ \hline
\hline
\textbf{Beta} & $\displaystyle{\ln {B(a_2, b_2)} +  \frac{1}{\alpha - 1}\ln \frac{B(a_\alpha, b_\alpha)}{B\left(\alpha\left(a_1-1\right)+1, \alpha\left(b_1-1\right)+1\right)}} $\\\cline{2-2} 
& $a_\alpha \coloneqq \alpha a_1 + (1-\alpha)a_2$, $\hspace{1cm} a_\alpha > 0$\\& $b_\alpha \coloneqq \alpha b_1 + (1-\alpha)b_2$, $\hspace{1cm} b_\alpha > 0$ \\ 
\hline
$\begin{array}{c}
\boldsymbol{\chi}\\ \textbf{ (scaled)}\end{array}$ & $\displaystyle{\frac{1}{2}\left( -\ln \frac{2\sigma_1^2}{\alpha}+k_2\ln\sigma_2^2\sigma^2_\alpha\right) +\ln\Gamma\left(\frac{k_2}{2}\right)}$\\
&$+ \displaystyle{ \frac{1}{1 - \alpha}\left( \frac{ \alpha k_1\ln\frac{\sigma^2_\alpha\sigma_1^2}{\alpha} }{2}-\ln\Gamma(\frac{k_\alpha}{2})+\ln\Gamma\left(\frac{\alpha(k_1-1)+1}{2}\right) \right)}$  \\\cline{2-2} 
	& $\sigma^{2}_\alpha \coloneqq \frac{\alpha}{\sigma_1^2}+ \frac{1-\alpha}{\sigma_2^2}$, $\hspace{1cm} \sigma^2_\alpha > 0$\\
	& $k_\alpha \coloneqq \alpha k_1+(1-\alpha)k_2$, $\hspace{1cm} k_\alpha > 0$\\
\hline 
$\begin{array}{c}
\boldsymbol{\chi}\\ \textbf{ (non-scaled)}\end{array}$ & $\frac{ -\ln 2\alpha}{2}+\ln\Gamma(\frac{k_2}{2})$\\
&$+ \displaystyle{\frac{1}{1-\alpha}\left(-\ln\Gamma(\frac{k_\alpha}{2})-\frac{\alpha k_1\ln \alpha}{2}+\ln\Gamma(\frac{\alpha(k_1-1)+1}{2})\right)}$  \\\cline{2-2} 
	& $k_\alpha \coloneqq \alpha k_1+(1-\alpha)k_2$, $\hspace{1cm} k_\alpha > 0$\\

\hline 
$\boldsymbol{\chi^2}$ & $\displaystyle{\frac{1}{1-\alpha}\left(-\ln\Gamma\left(\frac{\nu_\alpha}{2}\right)+\alpha \ln\Gamma\left(\frac{\nu_1}{2}\right)\right)}\displaystyle{+\ln\Gamma\left(\frac{\nu_2}{2}\right)} $  \\\cline{2-2} 
	& $\nu_\alpha \coloneqq \alpha\nu_1 + (1-\alpha)k$, $\hspace{1cm} \nu_\alpha > 0$\\

\hline

\textbf{Exponential} & $\displaystyle{\frac{1}{1-\alpha}\ln \frac{\lambda_1}{\alpha\lambda_\alpha} -\ln \lambda_2 }$  \\\cline{2-2} 
& $\lambda_\alpha \coloneqq \alpha\lambda_1 + (1-\alpha)\lambda_2$, $\hspace{1cm} \lambda_\alpha > 0$ \\

\hline 
\textbf{Gamma} & $\displaystyle{\ln \Gamma(k_2)+ k_2\ln\theta_2}$\\
& $+ \displaystyle{ \frac{1}{1-\alpha}\left( \ln\frac{\Gamma(k_1)}{\Gamma(k_\alpha)}-k_\alpha\ln\theta_\alpha-\alpha^2k_1\ln\theta_1 \right) } $ \\\cline{2-2} 
&  $\theta_\alpha \coloneqq \alpha\theta_1^{-1}+(1-\alpha)\theta_2^{-1}$, \ $k_\alpha \coloneqq \alpha k_1 + (1-\alpha)k_2$, $\hspace{1cm} \theta_\alpha > 0$  \\ 
\hline 

$\begin{array}{c}
\textbf{Gaussian}\\ \textbf{ (Univariate)}\end{array}$
 & $\displaystyle{\frac{1}{2}\left(\ln (2\pi\sigma_2^2) + \frac{(\mu_1 - \mu_2)^2}{ (\sigma^2)_\alpha}  +  \frac{1}{1-\alpha}\ln \left( \frac{\alpha\sigma_2^2}{(\sigma^2)_\alpha}\right)\right)} $\\\cline{2-2} 
& $(\sigma^2)_\alpha \coloneqq \alpha\sigma_2^2 + (1-\alpha)\sigma_1^2$, $\hspace{1cm} (\sigma^2)_\alpha > 0$ \\
\hline
$\begin{array}{c}
\textbf{Gaussian}\\ \textbf{ (Multivariate)}\end{array}$
 & $\displaystyle{\frac{1}{2-2\alpha}\left(-\ln|\alpha|+\ln|A||\Sigma_1|+d \right) +\frac{1}{2}\ln\frac{(2\pi)^n|\Sigma_1|^2}{|\Sigma_2|}} $\\\cline{2-2} 
& $A \coloneqq \alpha \Sigma_1^{-1}+(1-\alpha)\Sigma_2^{-1}$, $\hspace{1cm} A \succ 0$ \\
&$\displaystyle{d\coloneqq(\boldsymbol{\mu_1}-\boldsymbol{\mu_2})^T\Sigma_1A\Sigma_2(\boldsymbol{\mu_1}-\boldsymbol{\mu_2})}$\\
\hline
\pagebreak
\hline

$\begin{array}{c}
\textbf{Gumbel}\\
\boldsymbol{(\beta_1 = \beta_2 =\beta)}\
\end{array}$ & $\displaystyle{ \frac{\mu_2 + \alpha \mu_1}{\beta} +  \frac{1}{1-\alpha }\left(\ln \frac{\Gamma(2-\alpha)\eta_\alpha}{ \alpha\beta} +\frac{\mu_1}{\beta}\right)}$ \\\cline{2-2} 
& $\displaystyle{\eta_\alpha\coloneqq \alpha e^{-\mu_1/\beta}+(1-\alpha)e^{-\mu_2/\beta}}$, $\hspace{1cm} \eta_\alpha > 0$\\

\hline
%
 \textbf{Half-Normal}
 & $\displaystyle{\frac{1}{2}\left(\ln (\frac{\pi\sigma_2^2}{2}) +  \frac{1}{1-\alpha}\ln \left( \frac{\alpha\sigma_2^2}{(\sigma^2)_\alpha}\right)\right)} $\\\cline{2-2} 
& $(\sigma^2)_\alpha \coloneqq \alpha\sigma_2^2 + (1-\alpha)\sigma_1^2$ , $\hspace{1cm} (\sigma^2)_\alpha > 0$\\
\hline
$\begin{array}{c}
\textbf{Laplace}\\
\textbf{$(\mu_1=\mu_2=0)$}\\
\end{array}$
& $\displaystyle{ \frac{\ln b_\alpha+\ln\alpha b_1}{1-\alpha}+\ln2b_2}$
\\\cline{2-2} 
 & $\displaystyle{b_\alpha\coloneqq\frac{\alpha}{b_1}+ \frac{1-\alpha}{b_2}}$, $\hspace{1cm} b_\alpha > 0$\\
\hline
\hspace{-8pt}

\textbf{Maxwell Boltzmann}
&$\displaystyle{\frac{ -\ln 2+3\ln\sigma_2^2}{2}+\ln \frac{\alpha}{\sigma_1^2}}$  \\ & $+\displaystyle{\frac{1}{1-\alpha}\left(\frac{ 3}{2}\ln\frac{\sigma_\alpha\sigma_1^2}{\alpha}-\alpha\ln\frac{\sqrt{\pi}}{2} +\ln\Gamma\left(\alpha+\frac{1}{2}\right)\right) }$\\\cline{2-2} 
& $\sigma^{2}_\alpha \coloneqq \frac{\alpha}{\sigma_1^2}+ \frac{1-\alpha}{\sigma_2^2}$, $\hspace{1cm} \sigma^2_\alpha > 0$ \\

\hline
$\begin{array}{c}
\textbf{Pareto}\\
\boldsymbol{(m_1 = m_2=m)}
\end{array}$ 
& $\displaystyle{\frac{1}{1-\alpha}\bigg(\ln\lambda_\alpha-\ln\left(1-\alpha(\lambda_1-1)\right)\bigg)-\ln\lambda_2m }$
 \\\cline{2-2} 
& $\lambda_\alpha \coloneqq \alpha\lambda_1+\left(1-\alpha\right)\lambda_2$, $\hspace{1cm} \lambda_\alpha > 0$ \\ 

\hline

\textbf{Rayleigh} & $\displaystyle{\frac{\ln \sigma_1^2(\sigma^2)_\alpha+\ln\alpha+\ln\Gamma(\frac{1-\alpha}{2})}{1-\alpha}+\ln 2\sigma_1^2}
$ \\\cline{2-2} 
& $(\sigma^2)_\alpha \coloneqq \alpha \sigma_1^{-2} +\frac{1}{2}\ln \frac{2\sigma_1^4 \sigma_2^4}{\alpha}$, $\hspace{1cm} (\sigma^2)_\alpha > 0$ \\ 
\hline 
\end{longtable}

\bigskip
\section{R\'{e}nyi and Natural R\'{e}nyi Differential Cross-Entropy Rates for Stationary Gaussian Processes} \label{sec4}
In \cite{cole} the R\'{e}nyi differential cross-entropy rate for stationary zero-mean Gaussian processes was derived. This, alongside with the Shannon and Natural R\'{e}nyi differential cross-entropy rates, are summarised in Table \ref{RencrosenTable}. Here, $f(\lambda)$ is the spectral density of the first zero-mean Gaussian process, $g(\lambda)$ is the spectral density of the second,
$$h(\lambda)=f(\lambda)+(\alpha-1)g(\lambda),$$ 
and 
$$j(\lambda)=\alpha f(\lambda)+(1-\alpha)g(\lambda).$$

\medskip

\begin{longtable}{lcc} 
\caption{Differential Cross-Entropy Rates for Stationary Zero-Mean Gaussian Sources}\label{RencrosenTable} 
\\ \toprule 
\hspace{-8pt}
Information Measure
& Rate & Constraint \\ \midrule
\endfirsthead
\bottomrule
\endlastfoot
\hspace{-8pt}
$\begin{array}{l}
\text{Shannon}\\
\text{Differential}\\
\text{Cross-Entropy} \end{array}$ 
& $\displaystyle{\frac{1}{2}\ln2\pi + \frac{1}{4\pi}\int_{0}^{2\pi} \Big[ \ln g(\lambda) +\frac{f(\lambda)}{g(\lambda)}\Big] \, d\lambda}$  &$g(\lambda) > 0$
\\ \midrule 
\hspace{-8pt}
$\begin{array}{l}
\text{Natural R\'{e}nyi}\\
\text{Differential}\\
\text{Cross-Entropy}
\end{array}$ 
&  \hspace{-12pt}$\displaystyle{\frac{1}{2}\ln 4\pi^2 \alpha^{\frac{1}{\alpha - 1}} +\frac{1}{4\pi(1 - \alpha)}\int_{0}^{2\pi} \ln \frac{j(\lambda)}{ {g(\lambda)}^{\alpha}} \, d\lambda}$&$\displaystyle{\frac{j(\lambda)}{g(\lambda)} > 0}$  \\ \midrule 
\hspace{-8pt}

$\begin{array}{l}
\text{R\'{e}nyi}\\
\text{Differential}\\
\text{Cross-Entropy}
\end{array}$ 
&  \hspace{-12pt}$\displaystyle{ \frac{\ln 2\pi}{2}+ \frac{1}{4\pi(1-\alpha)}\int_0^{2\pi} \left[(2-\alpha)\ln {g}(\lambda) -\ln {h}(\lambda) \right] d\lambda}$&$\displaystyle{\frac{g(\lambda)}{h(\lambda)} > 0}$

\end{longtable}

\medskip

\section{R\'{e}nyi and Natural R\'{e}nyi Cross-Entropy Rates for Markov Sources}
\label{sec5}
In \cite{cole}, the R\'{e}nyi cross-entropy rate between finite-state time-invariant Markov sources was established, using as in~\cite{rached2001renyi} tools from the theory of non-negative matrices and Perron-Frobenius theory (e.g., cf.\ \cite{seneta2006non, gallager1996}). This, alongside the Shannon and Natural R\'{e}nyi differential cross-entropy rates, are derived and summarised in Table~\ref{MarkovcrosenTable}. Here, $P$ and $Q$ are the $m \times m$ (stochastic) transition matrices associated with the first and second Markov sources, respectively, where both sources have common alphabet of size $m$. To allow any value of the R\'{e}nyi parameter $\alpha$ in $(0,1)\cup (1,\infty)$, we assume that the transition matrix $Q$ of the second Markov chain has positive entries ($Q>0$); however, the transition matrix $P$ of the first Markov chain is taken to be an {\em arbitrary} stochastic matrix. For simplicity, we assume that the initial distribution vectors, $p$ and $q$, of both Markov chains also have positive entries ($p>0$ and $q>0$).\footnote{This condition can be relaxed via the approach used to prove \cite[Theorem~1]{rached2001renyi}.} Moreover, $\pi_p^T$ denotes the stationary probability row vector associated with the first Markov chain and $\mathbf{1}$ is an $m$-dimensional column vector where each element equals one. Furthermore, $\odot$ denotes element-wise multiplication (i.e., the Hadamard product operation) and $\dot \ln$ is the element-wise natural logarithm. 

Finally, the definition of $\lambda(R) : \mathbb{R}^{m\times m} \mapsto \mathbb{R}$ for a matrix $R$ is more involved. If $R$ is irreducible, $\lambda(R)$ is its largest positive eigenvalue. Otherwise, rewriting $R$ in its canonical form as detailed in~\cite[Proposition~1]{rached2001renyi}, we have that $\lambda(R) = \max(\lambda^*, \lambda_*)$, where $\lambda^*$ is the maximum of all largest positive eigenvalues of (irreducible) sub-matrices of $R$ corresponding to self-communicating classes, and $\lambda_*$ is the maximum of all largest positive eigenvalues of sub-matrices of $R$ corresponding to classes reachable from an inessential class.

\medskip

\begin{longtable}{cc} 
\caption{Cross-Entropy Rates for Time-Invariant Markov Sources}\label{MarkovcrosenTable} 
\\ \toprule 
\hspace{-8pt}
Information Measure
& Rate \\ \midrule
\endfirsthead
\bottomrule
\endlastfoot
\hspace{-8pt}
$\begin{array}{l}
\text{Shannon}\\
\text{Cross-Entropy} \end{array}$ 
&  $\displaystyle{ -\pi_p^T \left(P\odot \dot\ln Q\right)\mathbf{1}}$
\\ \midrule 
\hspace{-8pt}
$\begin{array}{l}
\text{Natural R\'{e}nyi}\\
\text{Cross-Entropy}
\end{array}$ 
&  \hspace{-12pt}$\displaystyle{\frac{1}{\alpha-1}\ln\frac{\lambda(P^\alpha \odot Q^{1-\alpha})}{\lambda(P^\alpha)}}$  \\ \midrule 
\hspace{-8pt}

$\begin{array}{l}
\text{R\'{e}nyi}\\
\text{Cross-Entropy}
\end{array}$ 
&  \hspace{-12pt} $\displaystyle{\frac{1}{1-\alpha}\ln\lambda(P\odot Q^{\alpha-1})}$

\end{longtable}

\medskip
\section{Conclusion}
\label{sec6}
We have derived closed-form formulae for the  R\'{e}nyi and Natural R\'{e}nyi differential cross-entropies of commonly used distributions from the exponential family. This is of potential use to further studies in information theory and machine learning, particularly problems where deep neural networks, trained according to a Shannon cross-entropy loss function, can be improved via generalized R\'{e}nyi-type loss functions in virtue of the extra degree degree of freedom provided by the R\'{e}nyi ($\alpha$) parameter. In addition, we have provided formulae for the R\'{e}nyi and Natural R\'{e}nyi differential cross-entropy rates for stationary zero-mean Gaussian processes and expressions for the cross-entropy rates for Markov sources. Further work includes expanding the present collection by considering distributions such as Levy or Weibull and investigating cross-entropy measures based on the $f$-divergence \cite{csi,csiszar67,silvey}, starting with Arimoto's divergence~\cite{liese}.

\medskip
\section*{Acknowledgements}
This work was supported in part by Natural Sciences and Engineering Research Council (NSERC) of Canada.


\section*{Appendix A: Distributions listed in Tables \ref{RenDivTable} and \ref{RenNatTable}}\label{app-b}
\begin{longtable}{cc} 
\hline
\textbf{Name} & \textbf{PDF} $f(x)$ \\
\textit{($\Theta$)} & (Support) \\\hline
\hline
\textbf{Beta} & $\displaystyle{{B(a, b)}x^{a-1}(1-x)^{b-1} } $\\
\textit{($a>0$, $b>0$)} & $\mathbb{S}=(0,1)$\\
\hline
\pagebreak
\hline
$\begin{array}{c}
\boldsymbol{\chi}\\ \textbf{ (scaled)}\end{array}$ & $\displaystyle{ \frac{2^{1 - k/2} x^{k - 1} e^{-x^2/2\sigma^2}}{\sigma^{k} \Gamma\left(\frac{k}{2}\right)}}$\\
\textit{($k >0$, $\sigma>0$)} & $\mathbb{S}=\mathbb{R^+}$\\
\hline 
$\begin{array}{c}
\boldsymbol{\chi}\\ \textbf{ (non-scaled)}\end{array}$ & $\displaystyle{ \frac{2^{1 - k/2} x^{k - 1} e^{-x^2/2}}{ \Gamma\left(\frac{k}{2}\right)}}$\\
\textit{($k >0$)} & $\mathbb{S}=\mathbb{R^+}$\\
\hline 
$\boldsymbol{\chi^2}$ & $\displaystyle{ \dfrac{1}{2^{\frac \nu 2} \Gamma\left(\frac \nu 2 \right)}x^{\frac \nu 2 -1}e^{-\frac x 2}}$\\
\textit{($\nu >0$)} & $\mathbb{S}=\mathbb{R^+}$\\
\hline 
\textbf{Exponential} & $\displaystyle{\lambda e^{-\lambda x}}$  \\
\textit{($\lambda>0$)} & $\mathbb{S}=\mathbb{R^+}$\\
\hline 
\textbf{Gamma} & $\displaystyle{\dfrac{1}{\theta^k\Gamma\left(k\right)}x^{k -1}e^{-\frac k \theta}}$\\
\textit{($k>0$, $\theta>0$)} & $\mathbb{S}=\mathbb{R^+}$\\
\hline 
$\begin{array}{c}
\textbf{Gaussian}\\ \textbf{ (univariate)}\end{array}$
 & $\displaystyle{\frac{1}{ \sqrt{2\pi\sigma^2} } e^{-\frac{1}{2}\left(\frac{x-\mu}{\sigma}\right)^2}} $\\
\textit{($\mu$, $\sigma^2>0$)} & $\mathbb{S}=\mathbb{R}$\\
\hline

$\begin{array}{c}
\textbf{Gaussian}\\ \textbf{ (multivariate)}\end{array}$
 & $\displaystyle{\frac{1}{ \sqrt{(2\pi)^n|\Sigma|} } e^{-\frac{1}{2}\left(\boldsymbol{x}-\boldsymbol{\mu}\right)^T\Sigma^{-1}\left(\boldsymbol{x}-\boldsymbol{\mu}\right)}} $\\
\textit{($\boldsymbol{\mu} \in \mathbb{R}^n$, $\Sigma^2 \in \mathbb{R}^{n\times n}\succ0$)} & $\mathbb{S}=\mathbb{R}^n$\\
\hline
 \textbf{Half-Normal}
 & $\displaystyle{\sqrt{\frac{2}{ \pi\sigma^2} } e^{-\frac{1}{2}\left(\frac{x}{\sigma}\right)^2}} $\\
\textit{($\sigma^2>0$)} & $\mathbb{S}=\mathbb{R^+}$\\
\hline
 \textbf{Gumbel}
 & $\displaystyle{ \frac{1}{ \beta}\exp-\left(\frac{x-\mu}{\beta} + e^{-\frac{x-\mu}{\beta}}\right)} $\\
\textit{($\mu$, $\beta>0$)} & $\mathbb{S}=\mathbb{R}$\\
\hline
 \textbf{Pareto}
 & $\displaystyle{am^ax^{-(1+m)}} $\\
\textit{($m > 0$, $a>0$)} & $\mathbb{S}+(m,\infty)$\\
\hline
 \textbf{Maxwell Boltzmann}
 & $\displaystyle{\frac{2x^2}{ \sqrt{\pi\sigma^6} } e^{-\frac{1}{2}\left(\frac{x}{\sigma}\right)^2}} $\\
\textit{($\sigma>0$)} & $\mathbb{S}=\mathbb{R^+}$\\
\hline
 \textbf{Rayleigh}
 & $\displaystyle{\frac{x}{\sigma^2 } e^{-\frac{1}{2}\left(\frac{x}{\sigma}\right)^2}} $\\
\textit{($\sigma^2>0$)} & $\mathbb{S}=\mathbb{R}^+$\\
\hline
 \textbf{Laplace}
& $\displaystyle{\frac{1}{2b} e^{-\frac{|x-\mu|}{b}}}$
\\
\textit{($\mu$, $b^2>0$)} & $\mathbb{S}=\mathbb{R}$\\
\hline
\end{longtable}

\pagebreak

\smallskip\noindent
{\bf Notes}

\begin{itemize}
    \item ${\displaystyle \mathrm {B}(a,b)=\int _{0}^{1}t^{a-1}(1-t)^{b-1}\,dt}$ is the Beta function.
    \item ${\displaystyle \Gamma(z)=\int _{0}^{\infty}x^{z-1}e^{-x}\,dx}$ is the Gamma function.
\end{itemize}

\bibliographystyle{IEEEtran}
\bibliography{citations}

\end{document}